\documentclass[12pt]{iopart}

\usepackage{epsfig,amsfonts}

\def\be{\begin{equation}}
\def\ee{\end{equation}}

\def\bea{\begin{eqnarray}}
\def\eea{\end{eqnarray}}

\def\e{\epsilon}

\def\ben{\begin{enumerate}}
\def\een{\end{enumerate}}

\def\bea{\begin{eqnarray}}
\def\eea{\end{eqnarray}}

\begin{document}

\title[On Gaudin model's scalar products and form factors]{On the determinant representations of Gaudin models' scalar products and form factors }
\author{Alexandre Faribault and Dirk Schuricht}
\address{Institute for Theory of Statistical Physics, RWTH Aachen, 52056 Aachen, Germany}
\ead{faribault@physik.rwth-aachen.de}
\begin{abstract}

We propose alternative determinant representations of certain form factors and scalar products of states in rational Gaudin models realized in terms of compact spins. We use alternative pseudo-vacuums to write overlaps in terms of partition functions with domain wall boundary conditions. Contrarily to SlavnovÕs determinant formulas, this construction does not require that any of the involved states be solutions to the Bethe equations; a fact that could prove useful in certain non-equilibrium problems. Moreover, by using an atypical determinant representation of the partition functions, we propose expressions for the local spin raising and lowering operatorsÕ form factors which only depend on the eigenvalues of the conserved charges. These eigenvalues define eigenstates via solutions of a system of quadratic equations instead of the usual Bethe equations. Consequently, the current work allows important simplifications to numerical procedures addressing decoherence in Gaudin models.

\end{abstract}

\date{\today}

\section{Introduction}

Integrable models based on the generalized Gaudin algebra \cite{gaudin,ortiz} have, in recent years, found a large ensemble of physical applications ranging from the mesoscopic BCS model \cite{bcs1,bcs2,bcs3} to  the central spin Hamiltonian \cite{centralspin1,centralspin2,centralspin3,centralspin4,centralspin5,centralspin6,centralspin7,centralspin8,centralspin9,centralspin10,centralspin11,centralspin12,centralspin13} through a variety of cavity based constructions relevant for quantum computing proposals \cite{dicke1,dicke2,dicke3}. The fact that their integrability does not necessitate strong restrictions on the model's parameters also makes them a remarkable playground to study externally tunable physical systems.

The exact eigenstates of Gaudin models are obtainable by finding sets of complex parameters (rapidities) which are solutions to an ensemble of non-linear algebraic equations known collectively as Bethe equations. However, the efforts to numerically solve these equations in a systematic fashion have shown it to be a challenging task \cite{solving1,solving2,solving3,solving4}. Recently an important improvement \cite{baode1,baode2,buccheri} has been achieved by exploiting a non-trivial change of variables based on the correspondence between Bethe equations and ordinary differential equations \cite{dorey1,dorey2}. In doing so, one can rewrite the problem in terms of quadratic equations  depending on a new set of variables $\Lambda(\epsilon_i)$ which are directly related to the eigenvalues of the model's conserved charges.

Using Lagrange's polynomial basis it was possible to implement an approach allowing one to extract the rapidities from a given set of  $\Lambda(\epsilon_i)$ \cite{baode2}. In doing so, one could turn to Slavnov's determinant \cite{slavnov} in order to compute scalar products and local operator form factors which are the elementary building blocks needed to address physical quantities. However, this work also motivated the search for simple representations of these quantities expressed, not in terms of  the rapidities themselves, but directly in terms of the easily found $\Lambda(\epsilon_i)$ variables. The current paper addresses this question and proposes to do so by using a non-standard determinant expression for the partition function with domain wall boundary conditions. In conjunction with the existence of two distinct representations for the eigenstates we find single determinant expressions for overlaps and spin raising/lowering operators form factors.

The paper is organized as follows. In section \ref{aba} we briefly review the Algebraic Bethe Ansatz (ABA) as applied to Gaudin models. Putting the emphasis on the two possible quantization axis $\pm \hat{z}$, we find a simple transformation between two equivalent representations of any eigenstate of the system. In Section \ref{ovnon} we then derive a determinant expression for the partition function with domain wall boundary condition which is used in Section \ref{scalp} to write scalar products of Bethe states as simple determinants. Section \ref{ff} concentrates on deriving determinant expressions for the form factors of local spin operators. In Section \ref{app} we discuss two possible applications of the obtained results to non-equilibrium problems.

\section{Algebraic Bethe Ansatz}
\label{aba}

Let us first introduce the generalized Gaudin algebra defined by the operators $ \mathrm{S}^x(u), \mathrm{S}^y(u),\mathrm{S}^z(u)$ satisfying the commutation relations\cite{gaudin,ortiz}:

\bea
\left[\mathrm{S}^x(u),\mathrm{S}^y(v)\right] &=& i(Y(u,v) \mathrm{S}^z(u)- X(u,v)\mathrm{S}^z(v)),
\nonumber\\
\left[\mathrm{S}^y(u),\mathrm{S}^z(v)\right]  &=&i(Z(u,v) \mathrm{S}^x(u)- Y(u,v)\mathrm{S}^x(v)) ,
\nonumber\\
\left[\mathrm{S}^z(u),\mathrm{S}^x(v)\right] &=&i(X(u,v) \mathrm{S}^y(u)- Z(u,v)\mathrm{S}^y(v)) ,
\nonumber\\
\left[\mathrm{S}^\kappa(u),\mathrm{S}^\kappa(v)\right] &=& 0, \ \ \ \kappa=x,y,z,
\label{commut}
\eea

\noindent where $u,v \in \mathbb{C}$. In this paper, we will deal only with the rational family of Gaudin models for which

\bea
X(u,v)=Y(u,v) = Z(u,v) &=&  \frac{1}{u-v}.
\eea

For a given number of excitations $M$, the ABA allows one to find eigenstates of the transfer matrix $T(u)=\mathrm{S}^2(u)$ using the following construction

\bea
\left|\lambda_1 ... \lambda_M\right>\equiv \prod_{i=1}^M \mathrm{S}^+(\lambda_i) \left| 0 \right>.
\label{bethestate}
\eea

\noindent Here $\mathrm{S}^+(u) = \mathrm{S}^x(u)+ i  \mathrm{S}^y(u)$ are generalized creation operators parametrized by the complex variable $u$. The pseudovacuum $\left|0 \right>$ is defined as the lowest weight vector, i.e. $\mathrm{S}^-(u) \left|0 \right> = 0, $ $\forall \ u  \in \mathbb{C}$.

States of the form (\ref{bethestate}) become eigenstates of 
\bea
T(u)\equiv\mathrm{S}^2(u)=\frac{1}{2}\left(\mathrm{S}^+(u)\mathrm{S}^-(u)+\mathrm{S}^-(u)\mathrm{S}^+(u)+2\mathrm{S}^z(u)\mathrm{S}^z(u)\right)
\label{tmatrix}
\eea

\noindent provided the $M$ rapidities $\lambda_i$ are solution of a set of coupled non-linear algebraic equation: the Bethe equations. For rational models, these equations can be written, in general, as

\bea
F(\lambda_i)= \sum_{j=1 (\ne i)}^M \frac{1}{\lambda_i-\lambda_j},
\label{eq:RGeq}
\eea

\noindent with 
\bea
S^z(\lambda_i)\left| 0 \right> = F(\lambda_i) \left| 0 \right>
\eea

\noindent defining the lowest weight function $F(u)$. 

Since one can show that $\left[\mathrm{S}^2(u),\mathrm{S}^2(v)\right]=0$, the operator-valued residues $\left\{R_1 ... R_N\right\}$ of $\mathrm{S}^2(u)$ at its arbitrarily chosen poles $u\in\left\{\epsilon_{1},... \epsilon_N\right\}$

allows one to define a set of $N$ commuting hermitian operators $R_i$. These become constants of motion for any integrable Hamiltonian obtained through linear combinations using coefficients $\eta_i \in \mathbb{R}$:

\bea
H = \sum_{i=1}^N \eta_i R_i.
\eea

\subsection{Correspondence between pseudo-vacuums}

When dealing with Gaudin models realized in terms of operators bounded from above and below, we have the freedom of defining the ABA using either the $\pm \hat{z}$ quantization axis.  Including an external magnetic field $\frac{1}{g} \hat{z}$, the two realizations in terms of $N$ local $su(2)$ spin operators of lenght $|S_i|$ are given by:

\bea
\begin{array}{lll}
\left|0\right> =\left| \downarrow ... \downarrow \right>
& \ & \left|0\right> = \left| \uparrow ... \uparrow \right> \\
 \mathrm{S}^z(u) = \frac{1}{g} - \displaystyle \sum_{i=1}^N \frac{S^z_i}{u-\epsilon_i}\equiv A(u)& \ & \mathrm{S}^z(u) = -\frac{1}{g} + \displaystyle \sum_{i=1}^N \frac{S^z_i}{u-\epsilon_i}     \\
  \mathrm{S}^+(u) =\displaystyle  \sum_{i=1}^N \frac{S^+_i}{u-\epsilon_i}\equiv B(u)& \ & \mathrm{S}^+(u) =\displaystyle  \sum_{i=1}^N \frac{S^-_i}{u-\epsilon_i}  \\
  \mathrm{S}^-(u) = \displaystyle \sum_{i=1}^N \frac{S^-_i}{u-\epsilon_i} \equiv C(u)& \ & \mathrm{S}^-(u) = \displaystyle \sum_{i=1}^N \frac{S^+_i}{u-\epsilon_i}    
\end{array},
\eea

\noindent where $\uparrow(\downarrow)$ respectively represent  the highest (lowest) weight state for each local spin. Note in passing that this readily excludes any model containing bosonic degrees of freedom such as Jayne-Cummings-Dicke-like models. Nonetheless, for any realization in terms of  finite magnitude spins or pseudo-spins, both constructions are available. 

Defining $\Omega = \displaystyle\sum_{i=1}^N 2 |S_i|$, the generic states containing M excitations above the lowest weight pseudo-vacuum 
\bea
\left|\lambda_1 ... \lambda_M\right> &\equiv& \prod_{i=1}^M B(\lambda_i) \left| \downarrow ... \downarrow \right>
\nonumber\\
\left|\mu_1 ... \mu_{N-M}\right> &\equiv& \prod_{i=1}^{\Omega-M} C(\mu_i) \left| \uparrow ... \uparrow \right>
\label{twostates}
\eea

\noindent turn into eigenstates of the transfer matrix provided the rapidities $\lambda_i$ or $\mu_i $ satisfy the Bethe equations (\ref{eq:RGeq}):

\bea
F^\lambda(\lambda_i) = -\sum_{k=1}^N\frac{|S_k|}{\epsilon_k-\lambda_i} +\frac{1}{g}&=&\sum_{j=1 (\ne i)}^M \frac{1}{\lambda_i-\lambda_j}
\nonumber\\
F^\mu(\mu_i) = -\sum_{k=1}^N\frac{|S_k|}{\epsilon_k-\mu_i} -\frac{1}{g}&=& \sum_{j=1 (\ne i)}^{\Omega-M} \frac{1}{\mu_i-\mu_j},
\label{be}
\eea

\noindent while the eigenvalues of $S^2(u)$ are then given by

\bea
\tau^\lambda(u) = \left[F^\lambda(u)\right]^2 -\frac{d}{du} F^\lambda(u) - 2 \displaystyle \sum_{i=1}^M \frac{F^\lambda(u)}{u-\lambda_i}+\sum_{i=1}^M\frac{1}{u-\lambda_i}\left(\sum_{j=1(\ne i)}^M\frac{1}{u-\lambda_j}\right)
\nonumber\\
\tau^\mu(u) = \left[F^\mu(u)\right]^2 - \frac{d}{du} F^\mu(u) - 2 \displaystyle \sum_{i=1}^{\Omega-M} \frac{F^\mu(u)}{u-\mu_i}+\sum_{i=1}^{\Omega-M}\frac{1}{u-\mu_i}\left(\sum_{j=1(\ne i)}^{\Omega-M}\frac{1}{u-\mu_j}\right).
\nonumber\\
\eea

The poles of these eigenvalues at $u=\epsilon_j$ give the eigenvalues $r_i$ of the commuting operators $R_i$, which are themselves read off from the poles of the $S^2(u)$ operator. Specializing to the non-degenerate case $(\e_i\ne\e_j \ \forall  \ i\ne j)$, we find:

\bea
R^\lambda_i = -\frac{ 2S^z_i }{g}  + \sum_{j=1(\ne i)}^N\frac{2\vec{S}_i \cdot \vec{S}_j }{\epsilon_i-\epsilon_j}  \ &\to& \ \frac{r^\lambda_i}{|S_i|} = -\sum_{j=1}^{M}\frac{2}{\epsilon_i - \lambda_j} + \frac{2}{g}+ \sum_{j=1 (\ne i)}^N \frac{2|S_j|}{\epsilon_i-\epsilon_j}
\nonumber\\
R^\mu_i = -\frac{2 S^z_i }{g}  + \sum_{j=1(\ne i)}^N\frac{2\vec{S}_i \cdot \vec{S}_j }{\epsilon_i-\epsilon_j}   \ &\to& \ \frac{ r^\mu_i }{|S_i|} = -\sum_{j=1}^{\Omega-M}\frac{2}{\epsilon_i - \mu_j} - \frac{2}{g}+ \sum_{j=1(\ne i)}^N \frac{2|S_j|}{\epsilon_i-\epsilon_j}.
\nonumber\\
   \label{gaudinh}
\eea

Unsurprisingly, one has the same conserved charges $R^\lambda_i = R^\mu_i $. In order to find a transformation leading from one representation of a given eigenstate to its other representation, it is sufficient to insure that every eigenvalues $r_i$ are the same in both cases.  In doing so, one easily sees that the transformation

\bea
\Lambda^\mu(\epsilon_i) = \Lambda^\lambda(\epsilon_i) - \frac{2}{g}
\label{trans}
\eea

\noindent does give the correspondence between both representations of a given eigenstate. Here we introduced the variables 
 
 \bea
  \Lambda^\lambda(\epsilon_i) &=&\sum_{j=1}^{M}\frac{1}{\epsilon_i - \lambda_j}
 \nonumber\\
 \Lambda^\mu(\epsilon_i)  &=&\sum_{j=1}^{\Omega-M}\frac{1}{\epsilon_i- \mu_j},
 \label{Lambdadef}
 \eea
 
 \noindent which are directly related to the eigenvalues $r_i$ of the commuting Gaudin Hamiltonians $R_i$ (see (\ref{gaudinh})). 
 
 One should keep in mind that the transformation is exclusively valid for states which are solutions to the Bethe equations (eigenstates) and that, evidently, the two representations can still differ by a normalization factor. Moreover, one should note that $\Lambda(\epsilon_i)$ are sufficient to allow a direct construction of the eigenenergies of any integrable Hamiltonian of the form $H = \sum_{i=1}^{N} \eta_i R_i$ with $\eta_i \in \mathbb{R}$.

Working with the rapidities $\{\lambda_1 ... \lambda_M\}, \{\mu_1 ... \mu_{N-M}\}$, establishing a transformation between both representations would only be possible by solving a further set of non-linear equations whereas here, using the $ \Lambda(\epsilon_i)$'s, it is remarkably simple. 

\subsection{Bethe equations for $\Lambda(\epsilon_i)$}

As briefly mentioned in the introduction, the $\Lambda(\epsilon_i)$ variables provide an extremely useful representation of the eigenstates in the sense that they obey a  set of algebraic equations which is much simpler than the underlying Bethe equations obeyed by the rapidities $\lambda_i$. 

For simplicity, the remainder of this paper will focus on non-degenerate realizations in terms of spin $\frac{1}{2}$ operators ($|S_k| =  \frac{1}{2})$. It was shown \cite{babelon} and exploited numerically \cite{baode1,baode2} that, in this case, solutions to the system of $N$ quadratic equations:

\bea
\left[ \Lambda^\lambda(\e_j)\right]^2
&=&
 \sum_{i=1 (\ne j)}^N \frac{\Lambda^\lambda(\e_j)-\Lambda^\lambda(\epsilon_i)}{\e_j-\epsilon_i}
+ \frac{2}{g}\Lambda^\lambda(\e_j)
\nonumber\\
\left[ \Lambda^\mu(\e_j)\right]^2 &=&\sum_{i=1 (\ne j)}^N \frac{\Lambda^\mu(\e_j)-\Lambda^\mu(\epsilon_i)}{\e_j-\epsilon_i}
- \frac{2}{g}\Lambda^\mu(\e_j)
\label{quadeqs}
\eea

\noindent are in one to one correspondence to solutions of the Bethe equations (\ref{be}) via the definitions (\ref{Lambdadef}). It is a trivial matter to verify that transformation (\ref{trans}) is consistent with both versions of eq. (\ref{quadeqs}).

\section{Partition function}
\label{ovnon}

Due to the relative simplicity of solving eqs (\ref{quadeqs}), it becomes highly desirable to be able to access physical quantities in terms of simple expressions involving exclusively the $\Lambda(\epsilon_i)$ variables. While Slavnov determinants fulfill such a role in terms of the rapidities $\lambda_i$, in the rest of this paper we will derive determinant expressions for scalar products and form factors of local spin operators in terms of the  $\Lambda(\epsilon_i)$ variables.

The first step, carried out in this section, is to show that the overlap of a generic Bethe-like state (\ref{twostates}) with an "infinite magnetic field ($g=0$)" eigenstate ($\left|\epsilon_{i_1} ... \epsilon_{i_M}\right> \equiv \displaystyle\prod_{j=1}^M S^+_{i_j} \left|\downarrow ... \downarrow \right>$) is writable as:

\bea
\left<\epsilon_{i_1} ... \epsilon_{i_M}\right.\left|\lambda_1 ... \lambda_M\right> = \mathrm{Det} J
\nonumber\\
\nonumber\\
J_{ab} =
 \left\{ \begin{array}{cc}
\displaystyle\sum_{c=1 (\ne a)}^M  \frac{1}{\epsilon_{i_a} -\epsilon_{i_c} }-\Lambda(\epsilon_{i_a}) & a= b
\\
\frac{1}{\epsilon_{i_a} -\epsilon_{i_b} }
& a\ne b \end{array}\right. .
\label{detrep}
\eea

In order to show this, one can start from the explicit construction of the state $\left|\lambda_1 ... \lambda_M\right> $  (eq. (\ref{twostates})), which leads to the formal expression: 

\bea
\left<\epsilon_{i_1} ... \epsilon_{i_M}\right.\left|\lambda_1 ... \lambda_M\right> = \sum_{\left\{P\right\}} \prod_{i=1}^M \frac{1}{\lambda_i-\epsilon_{P_i}}.
\label{perm}
\eea

\noindent Here $\left\{P\right\}$ is the ensemble of possible permutations of the indices $\left\{i_1 ... i_M\right\}$ and $P_i$ denotes the $i^{\mathrm{th}}$ element of the given permutation. By isolating in (\ref{perm}) the terms which depend on $\lambda_M$, one finds that the overlaps obey the simple recursion relation

\bea
\left<\epsilon_{i_1} ... \epsilon_{i_M}\right.\left|\lambda_1 ... \lambda_M\right> = \sum_{j=1}^M \frac{1}{\lambda_M - \epsilon_{i_j}} \left<\epsilon_{i_1} ... \hat{\epsilon}_{i_j} ... \epsilon_{i_M}\right.\left|\lambda_1 ... \lambda_{M-1}\right>,
\label{rec}
\eea

\noindent where $\left|\epsilon_{i_1} ... \hat{\epsilon}_{i_j} ... \epsilon_{i_M}\right>$ is the state with $M-1$ excitations, for which $\epsilon_{i_j}$ has been removed from the ensemble $\left\{\epsilon_{i_1} ... \epsilon_{i_M} \right\}$. 

This is obviously a rational function of $\lambda_M$, which goes to zero when $\lambda_M \to \infty$ and has only simple poles at every $\lambda_M = \epsilon_{i_j}$. To show that it does obey the recursion relation, it is therefore sufficient to show that the proposed determinant representation (\ref{detrep}) has the same poles $\lambda_M = \epsilon_{i_j}$ and the same residues $ \left<\epsilon_{i_1} ... \hat{\epsilon}_{i_j} ... \epsilon_{i_M} \right.\left|\lambda_1 ... \lambda_{M-1}\right>$ at these poles.  

 The determinant in (\ref{detrep}) clearly only has single poles at $\lambda_M = \epsilon_{i_j}$. Indeed, the $\epsilon_{i_j}$ pole comes only from the diagonal element $J_{jj}$ which, via $-\Lambda(\epsilon_{i_j})$, contains the term $\frac{1}{\lambda_M-\epsilon_{i_j}}$. The residue is trivially given by the determinant of the minor obtained by removing line and column $j$ after taking its $\lambda_M \to \epsilon_{i_j}$ limit:

\bea
\lim_{\lambda_M\to \epsilon_{i_j}}(\lambda_M -  \epsilon_{i_j}) \mathrm{Det}  J = 
\mathrm{Det}  J^{\hat{j}} \eea
\noindent with
\bea
J^{\hat{j}}_{a,b} = 
\left\{
\begin{array}{cc}
 \displaystyle\sum_{c=1(\ne a)}^M  \frac{1}{\epsilon_{i_a}-\epsilon_{i_c}} - \sum_{k=1}^{M-1} \frac{1}{\epsilon_{i_a}-\lambda_k} - \frac{1}{\epsilon_{i_a}-\epsilon_{i_j}} &  a=b \   (a,b \ne j)  \\
  \frac{1}{\epsilon_{i_a}-\epsilon_{i_b}} &  a\ne b \    (a,b \ne j)
\end{array}
\right. .
\eea

The diagonal elements of this matrix evidently reduce to $\displaystyle\sum_{c=1 (\ne j) }^M \frac{1}{\epsilon_{i_a}-\epsilon_{i_c}} - \sum_{\alpha=1}^{M-1} \frac{1}{\epsilon_{i_a}-\lambda_\alpha}$ and therefore correspond to the representation (\ref{detrep}) of $\left<\epsilon_{i_1} ... \hat{\epsilon}_{i_j} ... \epsilon_{i_M}\right.\left|\lambda_1 ... \lambda_{M-1}\right> $ proving the determinant obeys the recursion relation (\ref{rec}). 

Verifying that, for a single rapidity $\lambda_1$, the projection $\left<\epsilon_{i_1}\right.\left|\lambda_1\right> = \frac{1}{\lambda_1-\epsilon_{i_1}}$  is indeed equivalent to the 1 by 1 version of the above determinant $(-\Lambda_{i_1} = - \frac{1}{\epsilon_{i_1}-\lambda_1})$ then completes the proof.

This construction is in fact nothing but the partition function with domain wall boundary conditions which one would obtain using a reduced model which contains only the $M$ spins excited in the left state, i.e. using operators $\tilde{B}(\lambda) = \sum_{j=1}^M \frac{S^+_{i_j}} {\lambda-\epsilon_{i_j}}$:

\bea
\left<\epsilon_{i_1} ... \epsilon_{i_M}\right.\left|\lambda_1 ... \lambda_M\right> =\left<\uparrow_{i_1}\uparrow_{i_2} ... \uparrow_{i_M}\right|\prod_{i=1}^M\tilde{B}(\lambda_i) \left|\downarrow_{i_1}\downarrow_{i_2} ... \downarrow_{i_M}\right> .
\eea

Expression (\ref{detrep}) can however be contrasted with the appropriate limit of the more frequently encountered Izergin \cite{izerginold,izergin,korepinbook} determinant representation of such a scalar product, i.e.:

\bea
  \left<\epsilon_{i_1} ... \epsilon_{i_M} \right.\left|\lambda_1 ... \lambda_M\right> = \frac{\displaystyle\prod_{j,k=1}^{M}(\lambda_j-\epsilon_{i_k})}{\displaystyle\prod_{i>j =1}^M(\lambda_i-\lambda_j)\prod_{j<k =1}^M(\epsilon_{i_j}-\epsilon_{i_k})} \mathrm{Det} K
\nonumber\\
 \ K_{ab} = \frac{1}{(\epsilon_{i_b}-\lambda_a)^2}.
\label{typical}
\eea

\noindent which is not simply writable in terms of $\Lambda(\epsilon_i)$. One should keep in mind that the determinant expression (\ref{detrep}) (just as (\ref{typical})) is valid for any set of complex parameters $\lambda_i$ and does not require them to be solution to the Bethe equations.

Finally, it is worth pointing out that due to the invariance under the exchange of  the sets $\left\{\epsilon_{i_1}  ... \epsilon_{i_M} \right\}$ and $\left\{\lambda_{1} ... \lambda_{M} \right\}$ (as evidenced by expansion (\ref{perm})), one could also write the projection in terms of the rapidities themselves as the determinant of the following alternative $M$ by $M$ matrix:

\bea
J_{ab} =
 \left\{ \begin{array}{cc}
\displaystyle-\sum_{c=1 (\ne a)}^M  \frac{1}{\lambda_{a} -\lambda_{c} }+\displaystyle \sum_{c=1}^M \frac{1}{\lambda_a-\epsilon_{i_c}} & a= b
\\
-\frac{1}{\lambda_{a} -\lambda_{b} }
& a\ne b \end{array}\right. .
\eea

\section{Scalar products}
\label{scalp}

The scalar product between two generic states (eq. \ref{twostates}) built out of the two different representations using respectively $M$ and $N-M$ rapidities is then writable as

\bea
\left<\mu'_1 ... \mu'_{N-M}\right.\left|\lambda_1 ... \lambda_M\right> &=& \left<\uparrow...\uparrow\right|\prod_{i=1}^{N-M} B(\mu'_i)\prod_{j=1}^{M} B(\lambda_i)\left|\downarrow...\downarrow\right>\nonumber\\
&\equiv& \left<\uparrow...\uparrow\right.\left|\nu_1... \nu_N\right> ,
\eea

\noindent where $\left\{\nu_1, ... \nu_N\right\} = \left\{\mu'_1 ... \mu'_{N-M}\right\}\cup\left\{\lambda_1 ... \lambda_M\right\}$ is the union of both sets of rapidities and has cardinality $N$. In doing so, we are once again dealing with a partition function with domain wall boundary conditions, this time using the full set of $N$ local spins. The results of the previous section are directly usable and lead to the determinant of the $N\times N$ matrix:

\bea
&&\left<\mu'_1 ... \mu'_{N-M}\right.\left|\lambda_1 ... \lambda_M\right> = \mathrm{Det} K
\nonumber\\
\nonumber\\
K_{ab} &=&
 \left\{ \begin{array}{cc}
\displaystyle\sum_{c=1 (\ne a)}^N  \frac{1}{\epsilon_{a} -\epsilon_{c} }-\Lambda^\nu(\epsilon_a) & a= b
\\
\frac{1}{\epsilon_{a} -\epsilon_{b} }
& a\ne b \end{array}\right. 
\nonumber\\ &=&  \left\{ \begin{array}{cc}
\displaystyle\sum_{c =1(\ne a)}^N  \frac{1}{\epsilon_{a} -\epsilon_{c} }-\Lambda^\lambda(\epsilon_a) -\Lambda^{\mu'}(\epsilon_a) & a= b
\\
\frac{1}{\epsilon_{a} -\epsilon_{b} }
& a\ne b \end{array}\right. 
\label{overlaps}
\eea

We note that for any ensemble of rapidities whose union has cardinality $\ne N$, both states would have different magnetizations and would therefore be orthogonal.

Contrarily to the traditional Slavnov determinant for $\left<\lambda'_1 ... \lambda'_{M}\right.\left|\lambda_1 ... \lambda_M\right>$ which is only valid when one of the two states is a solution to the Bethe equations, the current expression has no restriction on any of the two sets of rapidities. Provided the $\mu'$-state is an eigenstate, it corresponds to an alternative $\lambda'$-state using transformation (\ref{trans}) and, in this specific case, we have

\bea
\left<\lambda'_1 ... \lambda'_{M}\right.\left|\lambda_1 ... \lambda_M\right> \propto \left<\mu'_1 ... \mu'_{N-M}\right.\left|\lambda_1 ... \lambda_M\right> = \mathrm{Det} K
\nonumber\\
\nonumber\\
K_{ab}  =   \left\{ \begin{array}{cc}
\displaystyle\sum_{c=1 (\ne a)}^N  \frac{1}{\epsilon_{a} -\epsilon_{c} }-\Lambda^\lambda(\epsilon_a)-\Lambda^{\lambda'}(\epsilon_a)+\frac{2}{g} & a= b
\\
\frac{1}{\epsilon_{a} -\epsilon_{b} }
& a\ne b \end{array}\right. .
\label{overlapseigen}
\eea

While the issue of the normalization will be discussed in the next section, we showed that by mixing both representations one can write the scalar products of unnormalized states in terms of $\Lambda(\epsilon_i)$ variables.

\subsection{Normalization}

For any state which allows both representations $\left|\lambda_1 ... \lambda_{M}\right>$ or $\left|\mu_1 ... \mu_{N-M}\right>$, the actual norm of either representation expressed in terms of the $\Lambda(\epsilon_i)$ variables remains elusive. However, their scalar product $\left<\mu_1 ... \mu_{N-M}\right.\left|\lambda_1 ... \lambda_{M}\right>$ is straightforwardly writable as a determinant. Since both representations correspond to the same normalized state $\left|\lambda_1 ... \lambda_{M}\right>_{Norm} = \frac{1}{N_\mu}\left|\mu_1 ... \mu_{N-M}\right> = \frac{1}{N_\lambda}\left|\lambda_1 ... \lambda_{M}\right>$, the mixed representation allows us to write

\bea
N_\mu N_\lambda = \left<\uparrow ... \uparrow\right|	\displaystyle \prod_{i=1}^{N-M} B(\mu_i) \prod_{i=1}^{M} B(\lambda_i)\left| \downarrow ... \downarrow\right> = \mathrm{Det} G
\eea

\noindent with the $N$ by $N$ matrix given by

\bea
G_{ab} = \left\{ \begin{array}{c}
\displaystyle\sum_{c=1 (\ne a)}  \frac{1}{\epsilon_{a} -\epsilon_{c} }-\Lambda^\lambda(\epsilon_{a}) -\Lambda^\mu(\epsilon_{a}) 
\\
\frac{1}{\epsilon_{a} -\epsilon_{b} }
 \end{array}\right. .
 \eea
 
In the specific case of eigenstates of the system, the correspondence (\ref{trans}) allows us to write it as
 \bea
G_{ab} = \left\{ \begin{array}{cc}
\displaystyle\sum_{c=1 (\ne a)}  \frac{1}{\epsilon_{a} -\epsilon_{c} }- 2 \Lambda^\lambda(\epsilon_{a})+\frac{2}{g} & (a= b)
\\
\frac{1}{\epsilon_{a} -\epsilon_{b} }
& (a\ne b) \end{array}\right. .
\eea

Provided expressions for the form factors $\left<\mu'_1 ... \mu'_{N-M}\right|\mathcal{O}\left|\lambda_1 ... \lambda_{M}\right>$, this product is  sufficient to write the eigenbasis representation the $\mathcal{O}$ operator:

\bea
\mathcal{O} &=&\displaystyle \sum_{\big\{\lambda'_1 ... \lambda'_{M}\big\},\big\{\lambda_1 ... \lambda_{M}\big\}}\frac{\left|\lambda'_1 ... \lambda'_{M}\right>\left<\mu'_1 ... \mu'_{N-M}\right|\mathcal{O}\left|\lambda_1 ... \lambda_{M}\right> \left<\mu_1 ... \mu_{N-M}\right|
}{\left<\mu_1 ... \mu_{N-M}\right.\left|\lambda_1 ... \lambda_{M}\right>\left<\mu'_1 ... \mu'_{N-M}\right.\left|\lambda'_1 ... \lambda'_{M}\right>}.
\nonumber\\
\eea

\noindent Here, one should understand that the notation uses the following correspondence $\frac{1}{N_\mu}\left|\mu_1 ... \mu_{N-M}\right> = \frac{1}{N_\lambda}\left|\lambda_1 ... \lambda_{M}\right>$ and  $\frac{1}{N_{\mu'}}\left|\mu'_1 ... \mu'_{N-M}\right> = \frac{1}{N_{\lambda'}}\left|\lambda'_1 ... \lambda'_{M}\right>$ while the double sum covers twice a complete set of eigenstates.

For any state, be it an eigenstate or not, which is writable using both representations, expectation values of a given operator would also be normalizable by writing them as:

\bea
\left<\mathcal{O}\right>_{\lambda_1 ... \lambda_{M}} = \frac{\left<\mu_1 ... \mu_{N-M}\right|\mathcal{O}\left|\lambda_1 ... \lambda_{M}\right> }{\left<\mu_1 ... \mu_{N-M}\right.\left|\lambda_1 ... \lambda_{M}\right>}.
\label{expec}
\eea

Having even shown how to go from one to the other via the transformation (\ref{trans}), we know with certainty that both representations are available for eigenstates of the system. However, for a generic state built out of arbitrary rapidities $\{\lambda_1 ... \lambda_M\}$ it is not assuredly possible to build an equivalent $\{\mu_1 ... \mu_M\}$ representation. Still, in Section \ref{dynba} we discuss a possible scenario where, without being an eigenstate of any given static model, a physically relevant time-dependent state would be such that these two possible representations exist at any time making (\ref{expec}) a usable construction.

\section{Form factors}
\label{ff}

In this section we derive determinant representations for form factors of local spin operators.

\subsection{$S^\pm_i$ form factors}

The solution to the quantum inverse problem for the models considered here allows one to write local spin operators in a remarkably simple fashion. Indeed, local spin raising operators are simply given by:

\bea
S^+_i =  \lim_{\gamma \to \epsilon_i } (\gamma - \epsilon_i) B(\gamma).
\eea

This fact allows one to derive simple expressions for their form factors. Using the multi-representation construction, we obtain for the form factor between unnormalized states with M and M+1 up-spins:

\bea
&& \left<\mu'_1 ... \mu'_{N-M-1}\right| S^+_i \left|\lambda_1 ... \lambda_{M}\right> =\left<\lambda_1 ... \lambda_{M}\right| S^-_i\left|\mu'_1 ... \mu'_{N-M-1}\right>^* \nonumber\\&& = \lim_{\gamma \to \epsilon_i } (\gamma - \epsilon_i) 
\left<\uparrow...\uparrow\right| \left(\prod_{i=1}^{N-M-1} B(\mu'_i) \right)  B(\gamma)  \left(\prod_{i=1}^{M} B(\lambda_i) \right)\left| \downarrow,...,\downarrow\right> \nonumber\\
&& =  \lim_{\gamma \to \epsilon_i } (\gamma - \epsilon_i) \ \mathrm{det} J,
\eea

\noindent where the matrix $J$ is given by eq (\ref{overlaps}) with the values of $\Lambda^\nu(\epsilon_a)$  obtained for the ensemble $\left\{\mu'_1 ... \mu'_{N-M-1},\gamma , \lambda_1... \lambda_M\right\}$. The determinant has a single pole at $\gamma = \epsilon_i$  and consequently, since

\bea
\lim_{\gamma \to \epsilon_i} \Lambda^\nu(\epsilon_{j\ne i}) = \Lambda^{\mu'}(\epsilon_j)+  \Lambda^{\lambda}(\epsilon_j) + \frac{1}{\epsilon_j - \epsilon_i} 
\nonumber\\
\lim_{\gamma \to \epsilon_i} (\gamma - \epsilon_i) \Lambda^\nu(\epsilon_i) = -1,
\eea

\noindent the resulting form factor is simply given by the determinant of the $(N-1) \times (N-1)$ matrix:
\bea
 \left<\mu'_1 ... \mu'_{N-M-1}\right| S^+_i \left|\lambda_1 ... \lambda_{M}\right>= \mathrm{det} J'
\nonumber\\ 
J'_{ab} =
 \left\{ \begin{array}{cc}
\displaystyle\sum_{c=1 (\ne a,i)}^N  \frac{1}{\epsilon_{a} -\epsilon_{c} }- \Lambda^{\mu'}(\epsilon_a) -  \Lambda^{\lambda}(\epsilon_a)  & a= b \ ( \ne i)
\\
\frac{1}{\epsilon_{a} -\epsilon_{b} } & a\ne b ÊÊ\ ( \ne i) \end{array} \right. .
 \eea

\noindent which excludes $\epsilon_i$ from the sums as well as line and column $i$.

\subsection{$S^z_i$ form factors}

The $S^z_i$ form factors are obtainable in a similar fashion except for the fact that one needs to explicitly use commutation relations of $A(u)$ and $B(u)$ operators. The inverse problem gives us

\bea
S^z_i = -\lim_{\gamma \to \epsilon_i } (\gamma - \epsilon_i) A(\gamma),
\eea

\noindent and therefore 

\bea
&&\left<\mu'_1 ... \mu'_{N-M}\right|S^z_i\left|\lambda_1 ... \lambda_M\right> \nonumber\\ && =  -\lim_{\gamma \to \epsilon_i } (\gamma - \epsilon_i) \left<\uparrow,...,\uparrow\right|\prod_{i=1}^{N-M} B(\mu'_i)A(\gamma)\prod_{j=1}^{M} B(\lambda_i)\left|\downarrow,...,\downarrow\right>.\nonumber\\
\eea

Using the commutation relations (\ref{commut}), it is a straightforward exercise to commute the $A$ operator until it reaches the right and acts on the pseudo-vacuum $\left|\downarrow,...,\downarrow\right>$. In doing so, one obtains the following sum:

\bea
&&\left<\mu'_1 ... \mu'_{N-M}\right|S^z_i\left|\lambda_1 ... \lambda_M\right> \nonumber\\
&=& - \frac{1}{2}  \left<\mu'_1 ... \mu'_{N-M}\right.\left|\lambda_1 ... \lambda_M\right> + \sum_{j=1}^M \frac{1}{\epsilon_i-\lambda_j}   \left<\mu'_1 ... \mu'_{N-M}\right|S^+_i\left|\lambda_1 ...\hat{\lambda}_{j} ... \lambda_M\right>.
\nonumber\\
\label{szff}
\eea

\noindent where every term is writable as a determinant. However, we did not manage to reduce this sum to a single determinant. Such a feat is possible \cite{links} for $\left<\lambda'_1 ... \lambda'_{M}\right|S^z_i\left|\lambda_1 ... \lambda_M\right> $ using the Slavnov construction in terms of the rapidities since all determinants then differ by a single column. Consequently, it appears that the particular expression found here cannot be useful in any numerical application which involves the computation of a large number of $S^z$ form factors; even more so considering the fact that it would still require explicit knowledge of the rapidities $\lambda_j$. While obtaining rapidities from the set of $\Lambda(\epsilon_i)$ is possible following the procedure outlined in \cite{baode2}, having done so would clearly make the use a single Slavnov determinant a better suited approach to the computation of  the form factors.

Nonetheless,  this construction still has the advantage that, contrarily to Slavnov's formulas, it remains valid even when both $\{\mu'\}$ and $\{\lambda\}$ are not solutions to Bethe equations. In Section \ref{dynba}, we discuss a potential scenario in which one could explicitly exploit this fact.

\section{Applications}
\label{app}

\subsection{Non-equilibrium dynamics}

One of the central motivations behind this work was to numerically address the decoherence in the central spin model. It describes a central spin $\vec{S}_0$ coupled to an external magnetic field $B \hat{z}$ and interacting via non-uniform hyperfine couplings $A_j$ with a bath of $N$ spins $\vec{S}_j$. Its Hamiltonian is obtained using a single integral of motion $H= \frac{1}{2}R_0$ and using the correspondence $B = -\frac{1}{g}$, $\epsilon_0 = 0$ $A_j = -\frac{1}{\epsilon_j}$ which leads to:

\bea
H = B S^z_0 + \sum_{i=1}^N A_i \vec{S}_0\cdot\vec{S}_i.
\eea

In order to compute the non-equilibrium dynamics of a generic initial state writable as Bethe-like construction one can use the set of determinants proposed in this work and alleviate the necessity of explicitly finding rapidities $\lambda_i$ in order to describe the eigenstates. Starting from an initial condition given by a coherent superposition of the central spin and any arrangement of the bath spins with the spins $\{i_1 ... i_M\}$ pointing up and the rest pointing down:

\bea
\left|\psi(0)\right> &=& \alpha \left|\uparrow_0; \downarrow ... \uparrow_{i_1} ... \uparrow_{i_M} ... \downarrow\right> + \beta \left|\downarrow_0; \downarrow ... \uparrow_{i_1} ... \uparrow_{i_M} ... \downarrow\right> \nonumber\\&\equiv& \alpha \left|\epsilon_0;\epsilon_{i_1} ... \epsilon_{i_M}\right> + \beta\left|\epsilon_{i_1} ... \epsilon_{i_M} \right>,
\eea
\noindent one can write the coherence factor as:

\bea
\left<\psi(t)\right| S^+_0\left| \psi(t)\right> \nonumber\\ \ =\alpha\beta \sum_{n,m}  \frac{\left<\epsilon_0;\epsilon_{i_1} ... \epsilon_{i_M}\right. \left|\{\lambda\}_n\right>\left<\{\mu\}_n\right| S^+_0\left|\{\lambda\}_m\right>\left<\{\mu\}_m\right. \left|\epsilon_{i_1} ... \epsilon_{i_M} \right>}{\left<\{\mu\}_n\right.\left|\{\lambda\}_n\right>\left<\{\mu\}_m\right.\left|\{\lambda\}_m\right>}e^{i(\omega_n-\omega_m) t}
\nonumber\\
\label{cf}
\eea

\noindent where $m,n$ respectively cover the complete sets of $M$ and $M+1$ excitations eigenstates with energies $\omega_{m,n}$. In light of the work presented here it should be clear that the eigenenergies, the form factors and the overlaps of the initial condition with eigenstates are all writable exclusively in terms of $\Lambda(\epsilon_i)$ variables. The proposed expressions become particularly useful for intermediate system sizes such that the extra computational cost associated with $N$ by $N$ determinants (instead of $M$ by $M$ for Slavnov's formulas) outweighs the cost of extracting the rapidities $\lambda$ from the set of $\Lambda(\epsilon_i)$. 

The gain in computation speed allows one to compute a large enough number of contributions to use Monte Carlo sampling in order to evaluate the sums in Eq. (\ref{cf}). The interested reader should consult \cite{faribaultCS}, in which the central spin decoherence problem has been studied for a large range of external magnetic fields.

\subsection{Dynamical Bethe Ansatz}
\label{dynba}

Finally, considering that a dynamical Ansatz $\left|\lambda_1(t) ... \lambda_M(t)\right>$ can, in certain scenarios, describe exactly the non-equilibrium wavefunction for Gaudin models \cite{gritsev} , the ideas developed in this work could prove useful in this particular context. Indeed, when studying problems involving the time-evolution of the Hamiltonian by an arbitrary variation of the "magnetic field" $g(t)$, it is possible to write exactly the time-evolved wavefunction using a dynamical Ansatz \cite{gritsev} 
\bea
\left|\psi(t)\right> \propto \left|\lambda_1(t) ... \lambda_M(t)\right> \equiv \prod_{i=1}^M B(\lambda_i(t))\left|0\right>, 
\eea

\noindent where a model-dependent set of classical equations of motion is obeyed by $\lambda_i(t)$:

\bea
\frac{d \lambda_i(t)}{dt} &=& f^\lambda_i \left(\lambda_1(t) \  ...\  \lambda_M(t),g(t)\right).
\eea

For an initial state $\left|\lambda_1(0) ... \lambda_M(0)\right> $ which is also representable as  $\left|\mu_1(0) ... \mu_{N-M}(0)\right>$ using the alternative pseudo-vacuum one can derive a set of classical equations of motion for both representations. It is therefore possible to find, at all times, two representations of the true time-evolved wavefunction, i.e. $\left|\psi(t)\right> \propto \left|\lambda_1(t) ... \lambda_M(t)\right> \propto \left|\mu_1(t) ... \mu_{N-M}(t)\right>$. Since the time-dependent state is no longer writable as a solution to a static Bethe equation, Slavnov's determinant would not be available to compute expectation values. However equation (\ref{expec}) still provides the time evolution of the expectation value of observables:

\bea
\left<\psi(t)\right| S^{\pm , z}_i \left|\psi(t)\right> = \frac{\left<\mu_1(t) ... \mu_{N-M}(t) \right| S^{\pm , z}_i \left|\lambda_1(t) ... \lambda_{M}(t) \right>}{\left<\mu_1(t) ... \mu_{N-M}(t) \right. \left|\lambda_1(t) ... \lambda_{M}(t) \right>}
\eea

\noindent in terms of simple $N$ by $N$ determinants (or a sum of them for $S^z$).

We do not claim here any superiority of  the proposed $\Lambda(\epsilon_i)$-dependent determinants over the usual Izergin ones (\ref{typical}). We simply want to draw attention to the fact that, in this context, form factors can, in principle, be written as partition functions which provide simple formulas valid at any time.

\section{Conclusions}

In this work we studied Gaudin models realized in terms of spins of finite magnitude whose spectrum is bounded from above and below such that the Algebraic Bethe Ansatz can be carried out using two distinct quantization axes. We showed that the correspondence between both representations of its eigenstates is remarkably simple in terms of the set of variables $\Lambda(\epsilon_i)$ directly related to the eigenvalues of the conserved operators. We derive a determinant representation of domain wall boundary condition partition functions written in terms of the variables $\Lambda(\epsilon_i)$. By mixing the two possible representations it was then possible to write overlaps and local spin raising (lowering) form factors as such a partition function, making them writable in terms of the proposed determinant. Finally, we also point out how these ideas can find direct applications in the numerical treatment of certain out-of-equilibrium problems.

\ack 

This work was supported by the German Research Foundation (DFG) through the Emmy-Noether Program.

\end{document}